\documentclass{revtex4}
\usepackage{graphicx}
\begin{document}
\title{The Cascades Proposal for the Deep Underground Science and Engineering Laboratory}

\author{W. C. Haxton$^1$ and J. F. Wilkerson$^2$}

\affiliation{Institute for Nuclear Theory$^1$ and Center for Experimental Nuclear Physics and Astrophysics$^2$, \\
University of Washington, Seattle, WA 98195}

\email{haxton@phys.washington.edu,
jfw@u.washington.edu}

\begin{abstract}
One of the options for creating a Deep Underground Science and Engineering Laboratory (DUSEL)
is a site in the Mt. Stuart batholith, a granodiorite and tonalite rock mass in the Cascade 
mountain range  in
Washington State.   The batholith's 100-year history in hard-rock tunneling includes the 
construction of the longest and deepest tunnels in the U.S., the parallel Cascade and Pioneer
tunnels.   The laboratory plan would utilize these two tunnels to produce a laboratory that
has many desirable features, including dedicated, clean, horizontal access,  
container-module transport, and low operations costs.   Various aspects of the site help to
reduce geotechnical, environmental, and safety risks. 
\end{abstract}

\date{today}
\maketitle

\section{Introduction}
In 2003 a search was done over most of the western U.S. to identify sites suitable for 
a horizontal-access DUSEL, a laboratory that will host future neutrino, dark matter, geoscience, geomicrobiology, and engineering projects \cite{deep}.  The Mt. Stuart batholith, a granodiorite formation in the Cascade Range, was one of the identified sites.  Because of a combination of railroad and water projects, this
batholith has been frequently tunneled over the past 100 years. In the most famous of these
projects, the 1926-29 excavation of the parallel Cascade and Pioneer tunnels, a world record for the rate of advance through hard rock was established.
These remain the longest (12.5 km) and deepest (1040 m) transportation tunnels in the U.S.
DUSEL interest in the Mt. Stuart batholith increased in August 2005 when the owner of the 
Cascade and Pioneer tunnels, Burlington Northern \& Santa Fe, decided that
development of the Pioneer Tunnel for science would be compatible with
railroad operations in the Cascade Tunnel, and that such development might simplify
tunnel maintenance and improve safety.   Subsequent work on DUSEL-Cascades \cite{haxton} has focused on using the
Pioneer Tunnel as the portal to a three-level laboratory, a horizontal Level I
at 1040m that would be ideal for large-detector construction, an intermediate Level II (1760 m) that
would be coupled to Level I by two hoists, and a very deep Level III (2330 m) that would be developed
a decade or more in the future.

The site has a number of attributes important to a project like DUSEL:
\begin{itemize}
\item  The site has a low-elevation portal (685 m) and mild climate, and is located in
the western foothills of the Cascades.  Thus the site is accessible year around, with the
drive from the Seattle-Tacoma airport (1.5 hours) generally snow-free.
\item  All surface development would occur on land owned by BNSF, and all underground
development would occur in areas where BNSF owns the mineral rights.   Current
site use includes a range of railroad industrial activities very similar to those required
for DUSEL.
\item  The site is within a federal energy corridor.  Redundant power feeds to DUSEL  with automatic
pickup are possible because substations are located at both portals.  Power rates are 
very favorable, about \$0.016/kW-hr.  There is the potential to tie into ESnet, Pacific Wave,  and other international
high-speed data corridors, to facilitate remote visualization and detector operations.
\item  The site's location relative to potential neutrino beams will position the U.S. to
play an important role in long-baseline physics.  The baseline to
FermiLab, 2630 km, is very close to ideal for an on-axis superbeam experiment, as noted
in the S1 study \cite{deep}.   Its neutrino-factory baselines are unique, with``magic" (7500 km) baselines
to both KEK and CERN, and with a FermiLab baseline that will optimize the followup
CP-violation experiment.  It is the only practical site on the globe matching 
APS Multi-Divisional Neutrino Study \cite{multi} neutrino-factory baseline specifications for each of
the three high-energy accelerator laboratories.
\item  The site is adjacent to a major international cargo-container route: a significant
fraction of the
cargo entering the U.S. from Asia comes by ship to the Seattle and Tacoma ports, then by
rail through the Cascade Tunnel to Chicago and points east.  One of the driving principles
of the DUSEL-Cascades design is to tie the laboratory seamlessly to this transportation
corridor, so that groups from distant laboratories can construct large modules at home, then
efficiently transport the modules to DUSEL-Cascades.
\item  The site is near a major urban center with a concentration of high tech industries, such
as aerospace, biotechnology, micro-electronics, and communications.  There are six major research
institutions within 200 miles.
\item  Risk factors are low.  The site has a favorable construction history and the rock on Level I
is accessible.  The site is clean and can be developed without creating rock
piles or other environmental legacies.  The ownership is stable.  The site's parallel tunnels can
be exploited to minimize underground ES\&H hazards.  The site is private with a
relatively uncomplicated set of permitting issues.  The design minimizes long-term operations and experimental
costs.  The proximity to Puget Sound will be helpful to DUSEL recruiting, opening opportunities for 
academic, industrial, and outreach partnerships and providing employment opportunities for spouses.
\end{itemize}

\section{Geologic, environmental, and geotechnical characterization}
\subsection{Regional geologic context}
The Mt. Stuart batholith is a 600 km$^2$ granitic formation,
dominantly granodiorite and tonalite with schist wallrock, located in a convergent tectonic
margin.  The host rock, Chiwaukum schist and banded gneiss, was intruded by the
Mt. Stuart batholith and other plutons during regional metamorphism in the Late Cretaceous period,
about 93 My ago.   Geologic work in the area began 120 years ago and includes surface
mapping, borehole studies, and construction of nine significant tunnels.  

The portal is located in a region of moderate seismic activity, separated by approximately 70 km
from the nearest identified active crustal fault.  No earthquake over magnitude 4.0 has occurred within
35 km over recorded history.  The U.S. Geologic Survey Seismic Hazard Mapping Index
for the site is 15\% of g peak acceleration (10\% probability over 50 years).  The corresponding
1997 Uniform Building Code risk zone is 2B.  (The range is 1 to 5, with 1 the lowest.)
The portal location was selected by BNSF in 1927 to be free of rock falls and
avalanches.

\subsection{Environmental characterization}
An analysis of the rock shows that the principal minerals are plagioclase, quartz, biotite,
hornblende, and K-feldspar.  No asbestiform minerals were identified.  The principal health
hazard from the rock will be quartz dust produced in construction, which is readily controllable
by standard industry practices.

The Pioneer Tunnel serves as the drainage gallery for the tunnel complex.  Tests of the
drainage from the Pioneer Tunnel were done  by RETEX Group Inc., assisted by
collaboration scientists from Pacific Northwest National Laboratory and by BNSF.   Analysis
included pH (7.29), temperature (15.6 C), hardness (28.4 mg/L), and a variety of metals.
Sulfide content is very low, below analytical sensitivity.   All results met the applicable
EPA or Washington State chronic standards.

Subsurface radioactivity was determined by gamma counting and XRF as a function of depth,
by counting rock samples taken from a recent deep glacial cut into the batholith.
The activities for U ($\sim$ 0.5 ppm), Th ($\sim$ 0.7 ppm), and K$_2$O ($\sim$ 1.4\%) are
very low for granitic rock.

Subsurface temperatures have been mapped from the surface to 1040 m, and thermal gradients
for projections to greater depth determined from these data, with suitable corrections for the
effects of surface topography.  The Level I (1040 m) rock temperature is 21 C, and the
estimated Level II (1760 m) and Level III (2330 m) temperatures are 34 $\pm$ 3 C and
45 $\pm$ 4 C, respectively.

The site is free of spotted owl or other protected nesting areas.  The area's drainage
basin is the Tye River, but the portal is well above waterfalls and other natural barriers.
Thus there are no Chinook or bull trout impacts associated with water release.

\begin{figure}[h]
\begin{minipage}{18pc}
\includegraphics[width=18pc]{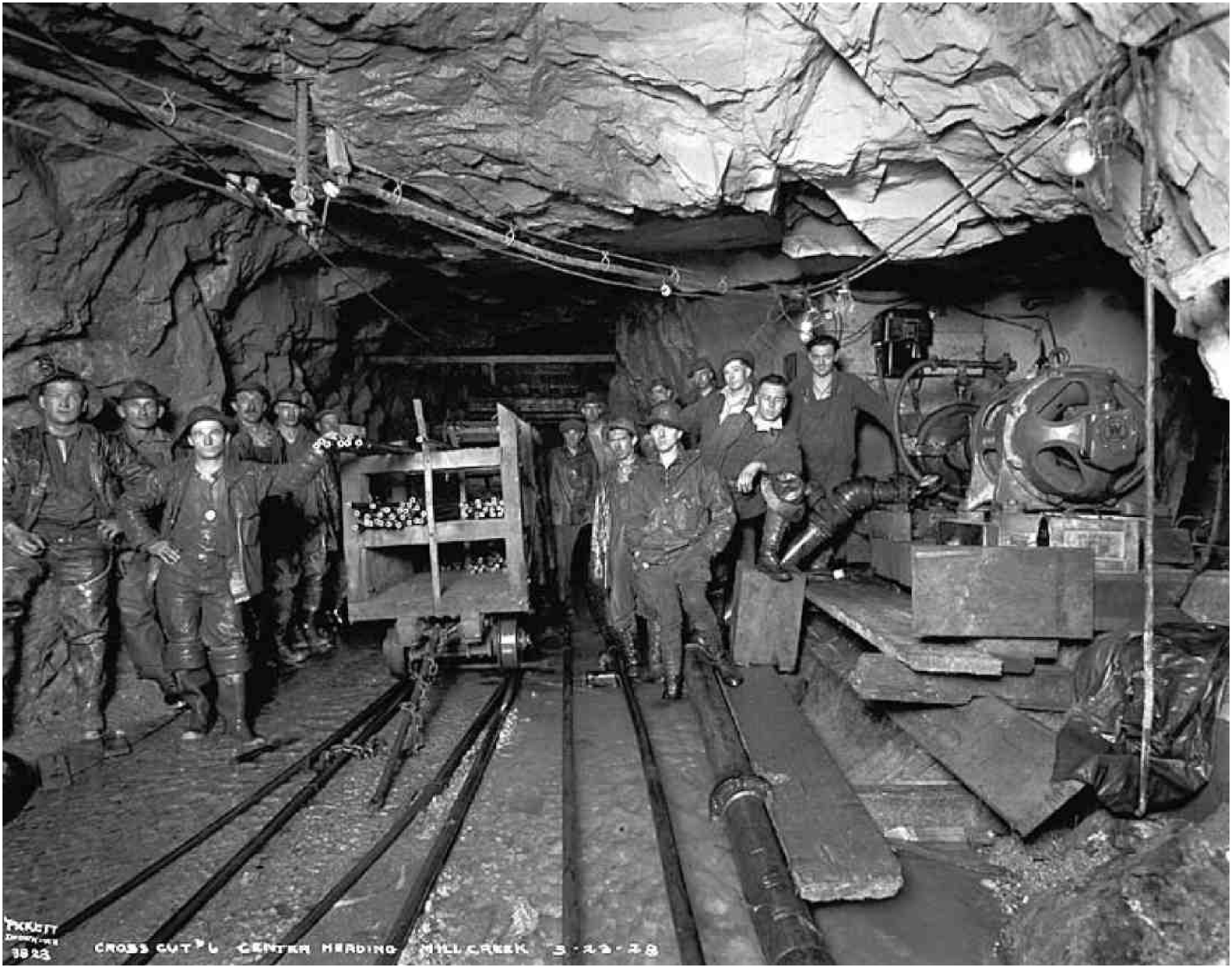}
\caption{\label{OT} Archival Pioneer Tunnel  photo showing
an enlarged-span equipment bay \cite{leepickett}.}
\end{minipage}\hspace{1pc}%
\begin{minipage}{18pc}
\includegraphics[width=18.7pc]{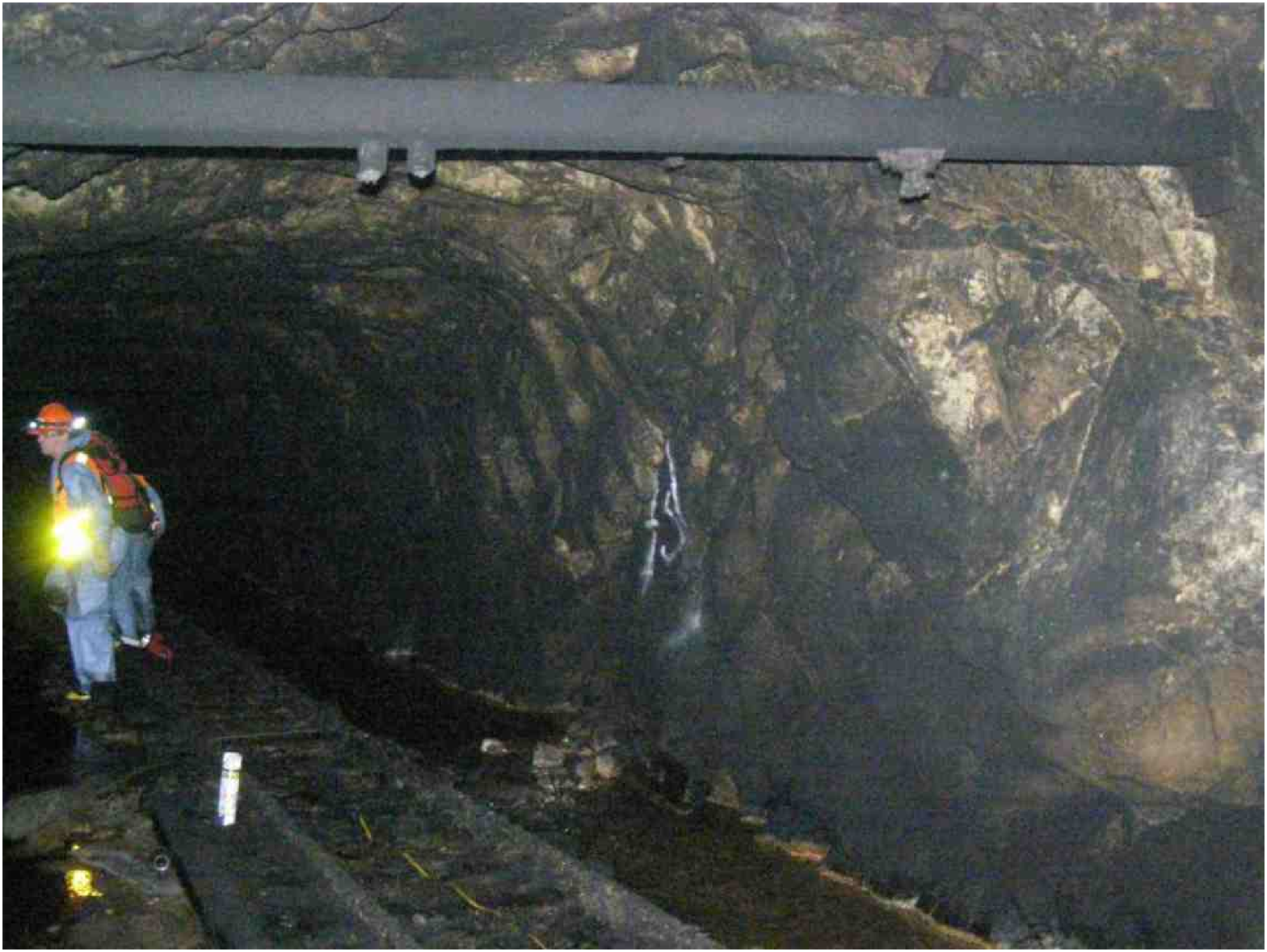}
\caption{\label{NT} Station 308 in the Pioneer Tunnel during recent reconnaissance \cite{red}.}
\end{minipage} 
\end{figure}

\subsection{Geotechnical characterization}
Part of the site's attractiveness is that it is effectively a greenfield -- the site does not
impose any significant constraints on DUSEL design -- without the usual geotechnical
uncertainties inherent in a greenfield.   The geotechnical database includes historical
records from construction and tunnel repairs, borehole studies conducted to determine
thermal gradients, extensive surface mappings (including a survey of surface outcroppings
commissioned by our collaboration in 2004), and surveys of the Pioneer Tunnel.  Because
the Pioneer Tunnel is largely unlined, it serves as a ``test adit," providing continuous access
to the rock along the DUSEL entrance tunnel and demonstrating the stability of openings
after 80 years.

The original construction records describe localized draining of newly opened rock.  Baxter
\cite{baxter} notes ``Practically all of the water encountered was ground-water filling fissures
in the rock.  These fissures were of all sizes... [and] acted in the same manner:  A maximum flow when first encountered;
a radical decrease in the flow in the next few days (sometimes hours), with a gradual decrease
over weeks and months, at times drying up entirely."   Level I rock is now dewatered.   
While fissures in granite tend to tighten with depth, new rock opened in the construction of Levels
II and III would likely exhibit similar transient drainage of the local rock mass: the contractor
would need to provide adequate pumping capacity during excavation.

The surface mappings of joints and fissures, compiled into stereonets \cite{web}, showed four major
joints sets, three of which are steeply dipping and one (the most common) shallow dipping.
Joint observations in the tunnel proved consistent with surface patterns: joint conditions
were found to improve with depth, with many joints/fractures either healed or discontinuous.

The most important database comes from recent geotechnical surveys along a 4500-ft section 
of the Pioneer Tunnel near maximum depth.  Tunnel walls were mapped, rock samples collected, and
laboratory tests performed for 45 tunnel stations located at 100-ft intervals.   Photographs from this reconnaissance (see Fig. \ref{NT}) and from
original construction (see Fig. \ref{OT}) are available \cite{leepickett,red}.
The resulting rock characterizations are summarized in Table \ref{toch}.  The results may be
conservative because laboratory tests were done with samples taken from exposed tunnel walls (rather then from 
corings).  Regardless, the rock quality is very high, comparable to that found at the Tochibora
Mine, a proposed HyperKamiokande site: the average RQD (Rock Quality Designation)
for the 45 stations is 94\% (excellent
rock) and the
average RMR (Rock Mass Rating) is 83 (Class I rock).  

\begin{center}
\begin{table}[h]
\caption{ \label{toch} Large Detector Sites}
\footnotesize\rm
\centering
\begin{tabular}{|l|c|c|c|}
\hline \hline
Site & Mozumi Mine & Tochibora Mine & Pioneer Tunnel\\
\hline
Principal rock type & Hornblende gneiss & Hornblende biotite gneiss& Granodiorite\\
Peak~~overburden (m) & 870 & 600-700 & $\sim$ 1000 \\
Density (g/cm$^3$) & 2.65 & 2.65 & 2.69-3.04 \\
Compressive strength (kpsi) & 15.2-17.4 & 21.7-36.3 & 16.4 $\rightarrow$ 35 \\
Discontinuity spacing (m) & 0.2-0.6 & 0.6-2.0 & 0.2-2.0 \\
~~~~~~~~~~~~~~~~~~condition & Slightly rough & Very rough & Rough, healed/discontinuous \\
~~~~~~~~~~~~~~~~~~orientation & Favorable & V. Favorable & Favorable/v. favorable \\
RQD & 78\% & 85\% & 94\% \\
RMR & 79 & 89 & 83 \\
Rock Mass Classification & II & I & I \\
\hline
\end{tabular}
\end{table}
\end{center}

The tunnel, unsupported over 87\% of its length, is in generally good condition after 80 years.
In a number of locations the Pioneer Tunnel was widened to spans of 5-10 m to accommodate
equipment or to form junctions with cross cuts (e.g., see Fig. \ref{OT}).  Standard tables of stand-up times indicate an RMR of at least 85 for any stable 80-year span above 5m, and 90+ for any span above 7.5m.  This empirical demonstration of rock competence is consistent with the laboratory results
described above.

\section{The Development Plan}
The development plan for DUSEL-Cascades is based on several principles:
\begin{itemize}
\item Careful engineering for efficient science, enhanced safety, and economical lifetime operations:  The design provides for dedicated
horizontal access with tracked transport and standardized turning radii to allow efficient, safe equipment transport; a ducted ventilation
system capable of isolating emergencies occurring in laboratory rooms or in Level II hallways;
 a dedicated exhaust path that will help keep
exit-ways open in an emergency; the capacity to sequester new construction from active laboratory
areas; and vertical alignment of the three laboratory levels for efficient connections and efficient exploitation of Level I's receiving and industrial support facilities.
\item Maximize cleanliness:  Environmental radioactivity is often the primary background in
underground experiments.  As a dedicated facility with all-electric operations, finished
hallways,  and ducted
ventilation, DUSEL-Cascades will be clean from the portal onward.  The design allows individual rooms (or
portions of rooms) to be held at Class 1-10, provides for a common 
Class 1000 clean area, a ``white area" at Class 100,000, and ``dirty-side" operations that will
in fact be quite clean, if compared to laboratories operating in shared mine facilities.
\item  Complementing SNOLab:  North America will soon have a very deep vertical-access
facility at SNOLab.  DUSEL-Cascades will complement SNOLab by providing 
a horizontal-access level at Kamioka depths, ideal for long-baseline megadetector construction, and an intermediate level at Frejus depths (4.05 km-water-equivalent) that can accommodate container-scale
experimental modules.  A third level at SNOlab depths is planned, when North America finds
itself in need of additional very deep space.
\item  Use of the site's unique transportation system:  The design envisions efficient transport
of 20-ft container modules from distant laboratories to any DUSEL level.  Level I is
effectively a siding on the BNSF railroad: loads can be moved by rail to the portal, mounted onto
laboratory flatcars, then tracked to Level I laboratory rooms.  Tracked container transport is
extended to deeper levels by a large-diameter, roll-in/roll-out hoist that effectively serves as
a track switching device between levels.
\item  International cooperation:  The combination of large detector capabilities, efficient cargo
container transport, complementarity to SNOLab, and magic baselines will
position DUSEL-Cascades to
play an important role in international science.  Seattle is the mainland U.S. port closest to Asia,
which currently lacks a deep laboratory.  We envision DUSEL-Cascades being operated 
cooperatively, part of a network of international laboratories (SNOLab, Kamioka, Gran Sasso,
etc.) that would collectively endeavor to accommodate new experiments in the most efficient way.

\end{itemize}
\begin{figure}
\begin{center}
\includegraphics[width=38pc]{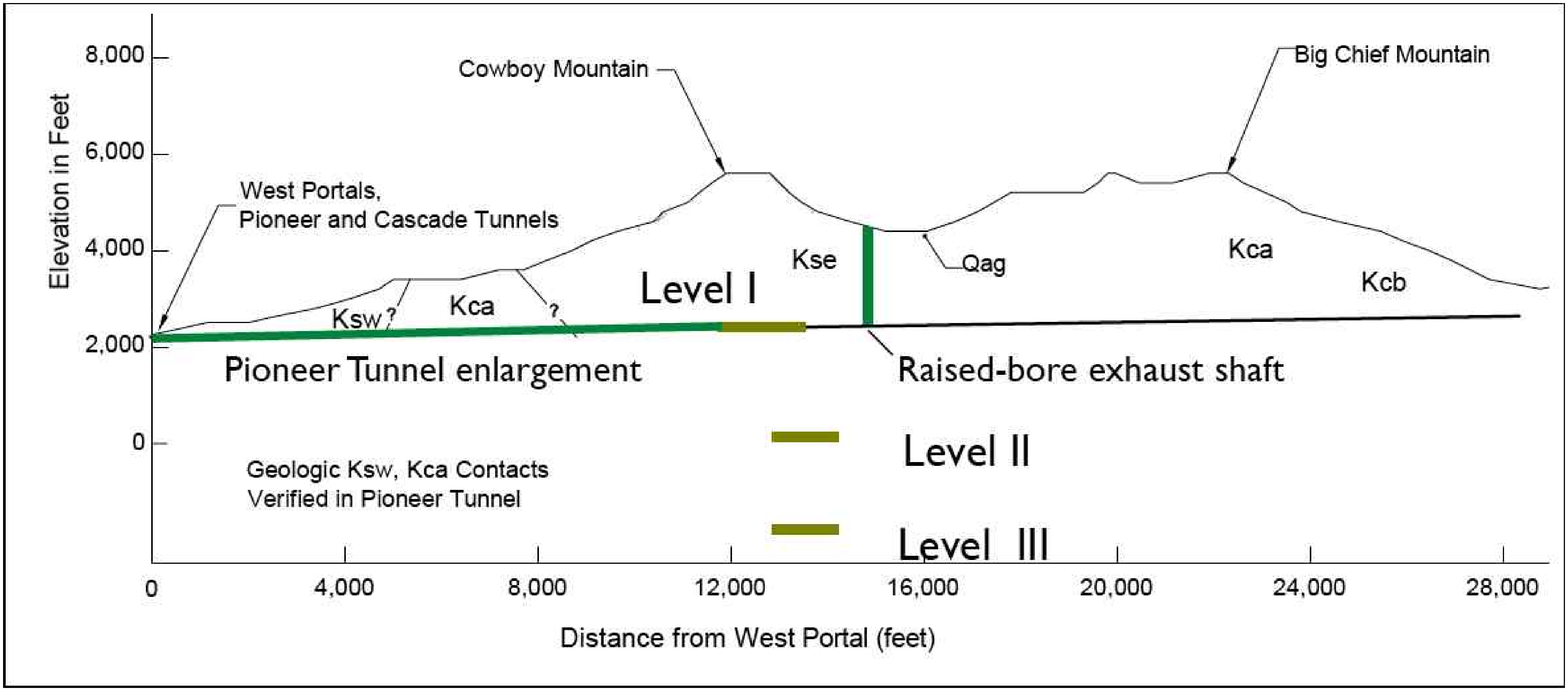}
\end{center}
\caption{\label{profile} Pioneer Tunnel topography and the arrangement of laboratory
levels, the entrance tunnel, and the exhaust shaft.}
\end{figure}

\begin{figure}[h]
\includegraphics[width=28pc]{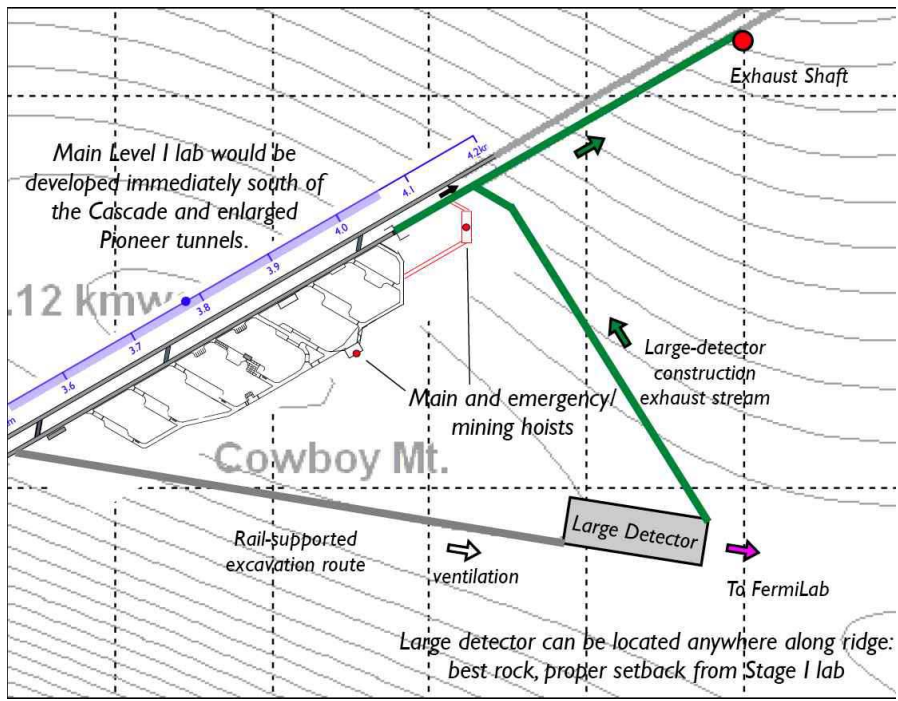}\hspace{1pc}
\begin{minipage}[b]{8pc}\caption{\label{levelI} The proposed arrangement of rooms and support facilities on Level I.
The site provides a large area, south of the tunnel, where a megadetector could be
constructed.  The layout shows how transportation to and exhaust from that construction
site could be routed to avoid interference with other Level I activities.}\end{minipage}
\end{figure}

Figure \ref{profile} shows the orientation of the three levels under Cowboy Mountain.  As
noted previously, significant work has been done to characterize rock properties on Level I.
The Pioneer Tunnel also provides a convenient platform for borehole studies of the deeper
rock that will house Levels II and III; both downhole and crosshole exploration programs can
be conducted at relatively low cost.  The deeper levels are connected to Level I -- the
industrial level providing access to the surface, power, drainage, ventilation, and receiving
services -- by a large-diameter container- and man-hoist, and by an emergency/mining hoist.
Levels II and III are designed so that new, specialized cavities can be constructed
``downstream" of other laboratory operations.  Thus these deeper levels can expand to
accommodate newly approved experiments.

Figure \ref{levelI} shows the plan for Level I, including a possible megadetector site.  The baseline design includes enlargement of
the Pioneer Tunnel (most likely by tunnel boring machine) to form the entrance hallway, 
with fully finished walls and a 
concrete tracked invert under which drainage would run; raise-bore construction of an exhaust shaft 
to the surface; and construction of a series of cross cuts/refuges between the Pioneer and Cascade
Tunnels, so that the latter can serve as a secondary escape route for the former.

Detailed discussions of this project can be found elsewhere \cite{web}.  This work was supported in
part by the Office of Nuclear Physics, U.S. Department of Energy.


\section{References}
\begin{thebibliography}{000}

\bibitem{deep} See the S1 report  {\it Deep Science}, http://www.dusel.org/

\bibitem{haxton} Haxton W C, Philpott K A, Holtz R, Long P, and Wilkerson J F 2007
{\it Nucl. Instruments and Methods in Physics Research A} {\bf 570} 414

\bibitem{multi} APS Multi-Divisional Neutrino Study, http://www.aps.org/neutrino/; also see
Huber P and Winter W 2003 {\it Phys. Rev. D} {\bf 68} 037301

\bibitem{leepickett} Photograph by Lee Pickett, University of Washington Digital Library.

\bibitem{red} Photograph by Red Robinson, Shannon \& Wilson, Inc.

\bibitem{baxter} Baxter J C 1932 {\it Transactions of the American Society of Civil
Engineers} {\bf 96} 950
 
\bibitem{web} See http://www.int.washington.edu/s3/

\end{thebibliography}
\end{document}